\documentclass[iop,apj]{emulateapj}
\usepackage{amsmath,amssymb,amstext}

\usepackage[breaklinks,colorlinks,citecolor=blue,linkcolor=magenta]{hyperref}

\usepackage{url}

\usepackage{natbib}
\newcommand{\sm}{M_\odot}
\newcommand{\sr}{R_\odot}
\newcommand{\RNum}[1]{\uppercase\expandafter{\romannumeral #1\relax}}

\begin{document}

\title{SN2002es-like supernovae from different viewing angles}

\author{Yi~Cao\altaffilmark{1}, S.~R.~Kulkarni\altaffilmark{1}, Avishay~Gal-Yam\altaffilmark{2},
S.~Papadogiannakis\altaffilmark{3}, P.~E.~Nugent\altaffilmark{4,5}, Frank~J.~Masci\altaffilmark{6}, 
Brian~D.~Bue\altaffilmark{7}}
\altaffiltext{1}{Astronomy Department, California Institute of Technology, Pasadena, CA 91125, USA}	
\altaffiltext{2}{Department of Particle Physics and Astrophysics, Weizmann Institute of Science, Rehovot 76100, Israel}	
\altaffiltext{3}{The Oskar Klein Centre, Department of Physics, Stockholm University, SE-106 91 Stockholm, Sweden}	
\altaffiltext{4}{Department of Astronomy, University of California, Berkeley, CA 94720-3411, USA}
\altaffiltext{5}{Lawrence Berkeley National Laboratory, 1 Cyclotron Road, MS 50B-4206, Berkeley, CA 94720, USA}
\altaffiltext{6}{Infrared Processing and Analysis Center, California Institute of Technology, Pasadena, CA 91125, USA}
\altaffiltext{7}{Jet Propulsion Laboratory, California Institute of Technology, Pasadena, CA 91125, USA}


\begin{abstract}
In this letter, we compare optical light curves of two SN2002es-like Type Ia supernovae, iPTF14atg and iPTF14dpk, 
from the intermediate Palomar Transient Factory. Although the two light curves resemble each other around and 
after maximum, they show distinct early-phase rise behavior in the \textit{r}-band. On the one hand, iPTF14atg
revealed a slow and steady rise which lasted for 22 days with a mean rise rate of $0.2\sim0.3\,\textrm{mag}\,\textrm{day}^{-1}$, 
before it reached the $R$-band peak ($-18.05$\,mag). On the other hand, iPTF14dpk rose rapidly to $-17$\,mag 
within a day of discovery with a rise rate $>1.8\,\textrm{mag}\,\textrm{day}^{-1}$, and then rose slowly to its peak
($-18.19$\,mag) with a rise rate similar to iPTF14atg. The apparent total rise time of iPTF14dpk is therefore only 
16 days. We show that emission from iPTF14atg before $-17$\,days with respect to its maximum can be entirely attributed to
radiation produced by collision between the SN and its companion star. Such emission is absent in iPTF14dpk probably because
of an unfavored viewing angle, provided that SN2002es-like events arise from the same progenitor channel. 
We further show that a SN2002es-like SN may experience a dark phase after the explosion but before its radioactively powered 
light curve becomes visible. This dark phase may be hidden by radiation from supernova-companion interaction. 
\end{abstract}


\section{Introduction}
\label{sec:14dpk:introduction}
Type \RNum{1}a supernovae (SNe \RNum{1}a) are explosions of carbon-oxygen white dwarfs (WDs). There are two leading 
scenarios about their origins. 
In the single degenerate (SD) channel, WDs accrete mass from non-degenerate companion stars and explode when 
their masses approach the Chandrasekhar mass limit. In the double degenerate (DD) channel, merging or 
collision of WD pairs in binary or triple systems triggers the SN explosions (See \citealt{mmn14} for a recent review). 

There are multiple subclasses of Type \RNum{1}a SNe: Branch-normal, SN1991T, super-Chandrasekhar, SN1991bg, 
\RNum{1}ax, SN2002es, etc.. The Type \RNum{1}ax \citep{fcc+13} and SN2002es-like \citep{glf+12} subclasses have low expansion velocities. 
Type \RNum{1}ax SNe lack \ion{Ti}{2} absorption lines in their spectra and favor young stellar populations, whereas 
SN2002es-like events show prominent \ion{Ti}{2} troughs and are associated with old
stellar population \citep{wkn+15}. Both low-velocity SNe are possibly pure deflagration of WDs. Their low expansion
velocities and large amount of intermediate mass elements are consistent with incomplete subsonic burning of carbon and oxygen \citep[e.g.,][]{fks+14}. 
Recent observations appear to find direct evidence for companion stars of two Type \RNum{1}ax SNe \citep{mjf+14,fmj+14}
and a SN2002es-like SN \citep{ckh+15}.

Following the discovery of a strong and declining UV pulse from a SN2002es-like supernova (SN) iPTF14atg within four days 
of explosion \citep{ckh+15}, it becomes particularly interesting to examine the early-phase light curve of other SN2002es-like events. 
If the UV pulse in iPTF14atg indeed arises from collision between the SN ejected material and a companion star, due to the viewing
angle effect, it should be invisible in most of other SN2002es-like events. 

Thanks to nightly-cadence surveys conducted as part of the intermediate Palomar Transient Factory (iPTF; \citealt{rkl+09,lkd+09}), 
we found two SN2002es-like events between 2013 and 2015, internally designated as iPTF14atg and iPTF14dpk. 
Both events have well-sampled optical light curves and spectroscopic coverage. 
In this letter, we present the early-phase rise behavior of these two events. 

The letter is organized as follows: In \S\ref{sec:14dpk:similarity} we establish their similarities to SN2002es.
In \S\ref{sec:14dpk:difference} we present the different rise behavior of the two events and seek an explanation. 
We conclude in \S\ref{sec:14dpk:conclusion}. 

\begin{deluxetable*}{cccccccc}
\centering
\tablecolumns{9}
\tablewidth{0pt}
\tablecaption{Two SN2002es-like Events in iPTF\label{tab:sample}}
\tablehead{
	\colhead{Name} & \colhead{Coordinate (J2000)} & \colhead{Redshift} & \colhead{$\mu$\tablenotemark{a}} & \colhead{Host Type} &\colhead{$E(B-V)$\tablenotemark{b}} 
	& \colhead{Peak MJD\tablenotemark{c}} & \colhead{Peak Mag.\tablenotemark{c}} 
	}
\startdata
	iPTF14atg & $12^{h}52^{m}44.84^{s}$~$+26^\circ28^\prime13.0^{\prime\prime}$ & $0.0213$ & $34.92$ & E-S0 & $0.011$ & $56802.1$ & $-18.05\pm0.02$ \\
	iPTF14dpk & $16^{h}45^{m}19.35^{s}$~$+40^\circ09^\prime41.3^{\prime\prime}$ & $0.0387$ & $36.23$ & Starburst & $0.012$ & $56878.1$ & $-18.19\pm0.02$
\enddata
\tablenotetext{a}{The distance moduli $\mu$ are calculated with $H=67.77\,\textrm{km}\,\textrm{s}^{-1}\,\textrm{Mpc}^{-1}$ \citep{planck13}. No redshift-independent 
distance measurement is available for the host galaxies of these events on the NASA/IPAC Extragalactic Database (NED). }
\tablenotetext{b}{The Galactic extinction map is given by \citet{sf11}.}
\tablenotetext{c}{The peak modified Julian dates (MJD) and magnitudes are measured from the PTF \textit{R}-band
light curves. The peak magnitudes do not include uncertainties from $\mu$. }
\end{deluxetable*}

\section{Similarity of the two events}
\label{sec:14dpk:similarity}
Observations of iPTF14atg and iPTF14dpk were undertaken in the Mould \textit{R} filter 
with the 48-inch Schmidt telescope at Palomar Observatory. Their light curves are produced
by the PTF-IPAC forced photometry pipeline (Masci et al. in prep.) which uses pre-SN 
reference images to subtract off the host galaxy light, and performs 
point-spread function photometry at the location of a transient on the difference images. 
All magnitudes in this paper are in the AB system and calibrated to the PTF-IPAC catalog 
\citep{ols+12}. 

The spectra of iPTF14atg were published in \citet{ckh+15}. 
The spectra of iPTF14dpk were obtained at $-10$ days with respect to its peak with the double 
spectrograph (DBSP; \citealt{dbsp}) on the 
Palomar 200-inch Hale telescope, and at $+20$ and $+50$ days with the Low-Resolution Imaging Spectrometer 
(LRIS; \citealt{lris}) on the Keck-\RNum{1} telescope at Mauna Kea. The spectra were reduced 
using standard IRAF/IDL routines. The spectra are made public through WISeREP \citep{yg12}.

In order to correct for the Galactic extinction, we use the \citet{f99} model assuming $R_V=3.1$.  
Regarding the host galaxy extinction, since the equivalent widths of \ion{Na}{1} D absorption lines 
are correlated with the local extinction \citep{ppb12}, the absence of \ion{Na}{1} D absorption 
at the host galaxy redshifts in the spectra suggests that
neither iPTF14atg nor iPTF14dpk is embedded in a dusty circumstellar medium. 
In fact, using the highest signal-to-noise ratio spectra of iPTF14atg and iPTF14dpk, assuming
the width of the \ion{Na}{1} D doublet is 200\,$\textrm{km}\,\textrm{s}^{-1}$, we derived 3$\sigma$
upper limits on the equivalent width of the doublet: $<0.20\,\textrm{\AA}$ for iPTF14atg and 
$<0.75\,\textrm{\AA}$ for iPTF14dpk. According to the empirical relation in \citet{ppb12}, 
these upper limits on the equivalent width corresponds to $E(B-V)<0.02$ and $E(B-V)<0.11$, 
respectively. In the following analysis, we do not correct for local extinction.

\begin{figure}
\centering
\includegraphics[width=0.495\textwidth]{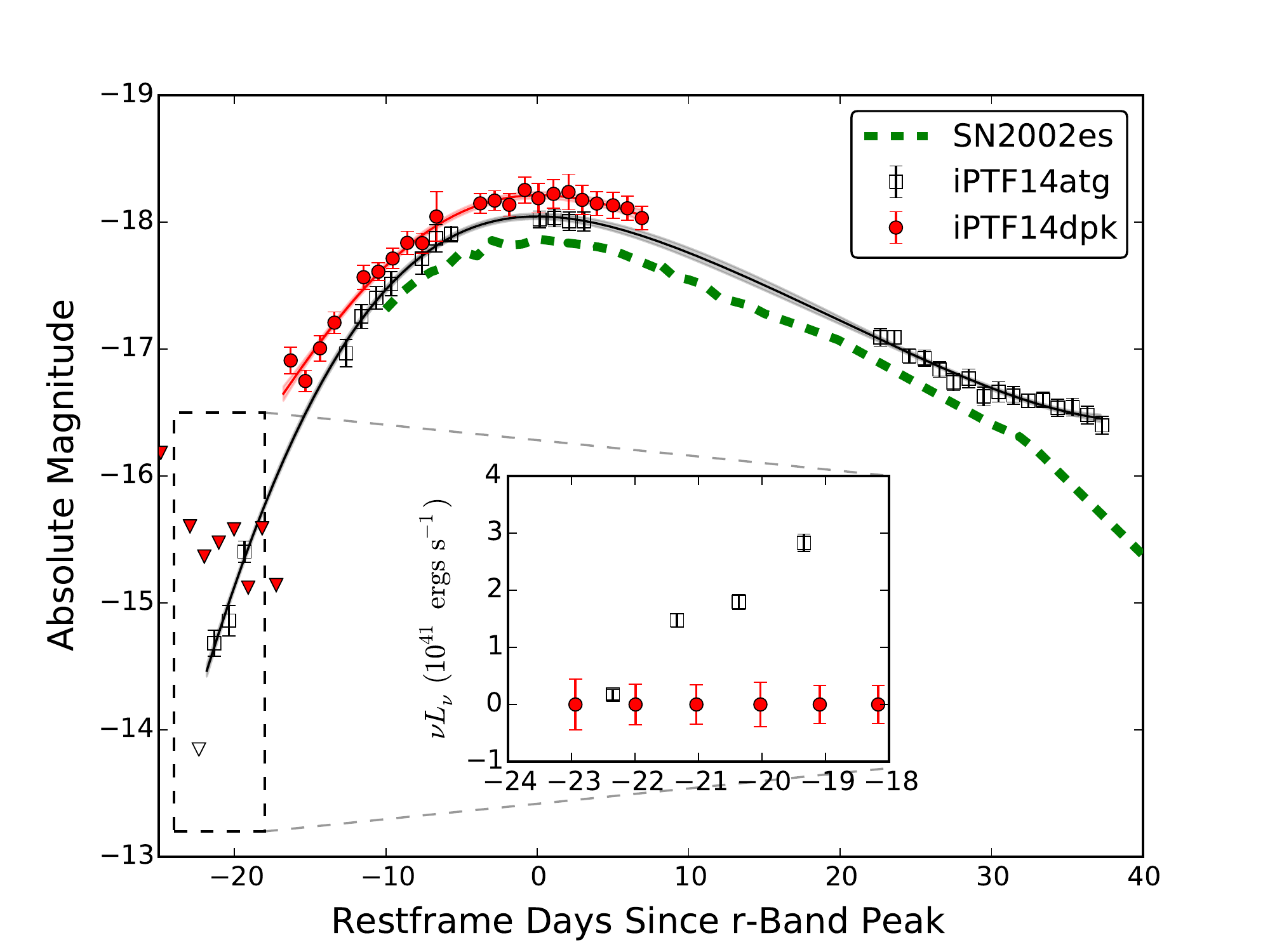}
\caption[Light curves of iPTF14atg and iPTF14dpk]{\textbf{Light curves of iPTF14atg and iPTF14dpk.}
Black empty squares are iPTF14atg and red filled stars denote iPTF14dpk. Detection upper limits (5-$\sigma$) of
pre-SN non-detections for iPTF14atg and iPTF14dpk are denoted by black empty and red filled downward triangles, respectively. 
The best interpolated light curves of iPTF14atg and iPTF14dpk with uncertainties from the Gaussian process regression are shown in red and black curves, respectively. For comparison, 
the green dashed curve shows the \textit{r}-band light curve of SN2002es. The inset zooms in on 
the early-phase light curves of iPTF14atg and iPTF14dpk in the $\nu L_\nu$ space. The error bars
represent 1$\sigma$ uncertainties of each measurements (instead of 5$\sigma$ upper limits in 
circumstances of null detection in the magnitude space). 
\label{fig:14dpk:lc}}
\end{figure}

\begin{figure}
\centering
\includegraphics[width=0.495\textwidth]{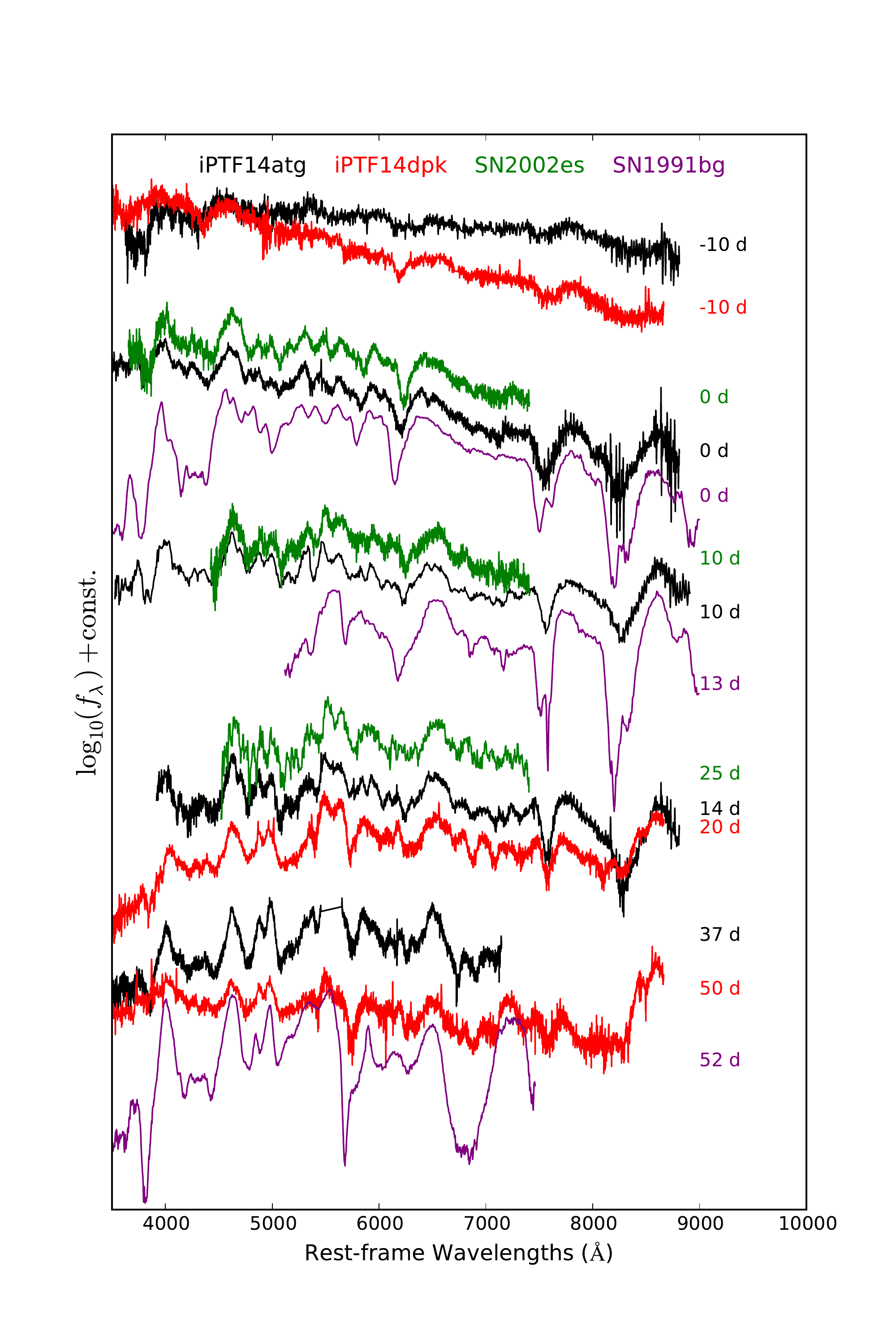}
\caption[Spectral sequences of iPTF14atg and iPTF14dpk]
{\textbf{Spectral sequences of iPTF14atg and iPTF14dpk.}
The iPTF14atg spectra are taken from \citet{ckh+15} and shown in black while the iPTF14dpk spectra are in red. 
For comparison, we also show spectra of SN2002es (green; \citealt{glf+12}) and SN1991bg (purple, \citealt{frb+92}). 
The phase of each spectrum is shown next to the spectrum. 
\label{fig:14dpk:spec}}
\end{figure}

The primary data of the two events are summarized in Table~\ref{tab:sample}. Their light curves 
and spectra are shown in Figures \ref{fig:14dpk:lc} and \ref{fig:14dpk:spec}, respectively.

Next, in order to determine their peak magnitudes and dates, we use Gaussian process regression to 
interpolate the light curves (Figure \ref{fig:14dpk:lc}). A squared exponential function is chosen as the 
autocorrelation function and 
the autocorrelation length is determined by the maximum likelihood estimation based on available data. 
Thanks to the almost nightly photometric sampling, the regression analysis determines both the peak dates and magnitudes
with small uncertainties, as shown in Table \ref{tab:sample}. 

We calculate \textit{k}-correction in the Mould \textit{R}-band for these two SNe using their spectra. We find that 
the \textit{k}-correction is $\lesssim0.2$\,mag. Hence, going forward, we neglect the \textit{k}-correction.

As shown in Figure \ref{fig:14dpk:lc}, apart from the first few days after explosions, the light curves of 
both iPTF14atg and iPTF14dpk are quite similar. Since the Mould \textit{R} filter is similar to the SDSS
\textit{r} filter \citep{oll+12}, light curves of both iPTF14atg and iPTF14dpk 
also resemble the \textit{r}-band light curve of SN2002es around maximum, though iPTF14atg does not 
show the fast decline seen in SN2002es after $t\simeq30\,\textrm{days}$. 
The peak magnitudes of iPTF14atg ($R=-18.05$\,mag), iPTF14dpk ($R=-18.19$\,mag) and SN2002es
($r=-18.35$\,mag) are also comparable within the slight filter difference.

As can be seen in Figure \ref{fig:14dpk:spec}, 
the spectra of iPTF14atg and iPTF14dpk around and after maxima also match those of SN2002es at similar phases. In particular, both 
iPTF14atg and iPTF14dpk show the low-velocity absorption lines and \ion{Ti}{2} troughs, hallmark of SN2002es-like SNe. Next, the spectra of iPTF14atg 
and iPTF14dpk taken at $-10$\,days share broad absorption features, although the continuum emission of iPTF14dpk appears 
bluer than that of iPTF14atg. This color difference might be due to residuals from imperfect subtraction of galaxy light in the spectra of iPTF14dpk, as
iPTF14dpk is located close to the center of its apparently blue and compact host galaxy in an interacting galaxy pair. 
Therefore, we do not think that the blueness of the iPTF14atg indicates significant difference from iPTF14dpk. 

Integrating the spectra and calibrating to the broadband photometry, we obtain an optical luminosity of $3\times10^{42}\,\textrm{ergs}\,\textrm{s}^{-1}$
for both SNe around maximum. Since SN radiation around maximum is concentrated in the optical, the optical luminosity is a good approximation to
the bolometric luminosity. In the case of iPTF14atg, the explosion date is tightly constrained by the SN-companion collision and 
therefore its rise time to the $R$-band maximum is $22$ days. The total $^{56}$Ni mass can be estimated from the
peak luminosity and the rise time via the following equation: 
\begin{equation}
L_{\textrm{max}}=\alpha S(t_R)\ ,
\end{equation}
where $S(t_R)$ is the instantaneous radioactive power at the light curve peak that can be expressed as
\begin{equation}
\frac{S(t)}{10^{43}\,\textrm{erg}\,\textrm{s}^{-1}} = \left[6.31\exp(-t/8.8)+1.43\exp(-t/111)\right] \frac{M_{\textrm{Ni}}}{\sm}\ ,
\end{equation}
where $t$ is in units of days. $\alpha$ is an efficiency factor
of order unity, depending on the distribution of $^{56}$Ni. We adopt a fiducial value of $\alpha=1.3$ here
following \citet{saa+12}. Then we estimate a total $^{56}$Ni mass of $0.14\sm$ for iPTF14atg. 

In contrast, as discussed in \S\ref{sec:14dpk:difference}, iPTF14dpk may experience a ``dark'' period after the explosion. Therefore its light curve 
only provides an upper limit on the actual explosion date, or equivalently a lower limit on the rise
time, and thus a lower limit of $0.11\sm$ on the $^{56}$Ni mass. 

In addition, we would like to clarify that the main difference of SN2002es-like events from ``classical'' 
subluminous SN1991bg-like events is that the former exhibit low velocity absorption features. 
As shown in Figure \ref{fig:14dpk:spec}, around maximum,
although the overall spectral shapes look similar among SN2002es-like events and SN1991bg, the velocities of the absorption lines, such as \ion{Si}{2},
are obviously lower in SN2002es-like events than in SN1991bg by several thousand $\textrm{km}\,\textrm{s}^{-1}$. At late time, SN2002es-like events develop multiple narrow 
features between 5000\,\AA\ and 7000\,\AA\ while these features are blended in the spectra of SN1991bg-like events \citep{jbc+06}.

\section{light curves at early phases}
\label{sec:14dpk:difference}

As noted in the previous section, iPTF14atg and iPTF14dpk exhibit different photometric 
behavior at early phases. In particular, 
iPTF14atg took more than a week to rise to $-17$\,mag with a mean rise rate of $0.2\sim0.3\,\textrm{mag}\,\textrm{day}^{-1}$. 
In contrast, iPTF14dpk shows a steep rise to $r=-16.9$\,mag within one day of 
its last non-detection of $r>-15.1$\,mag, indicating an initial rise rate $>1.8\,\textrm{mag}\,\textrm{day}^{-1}$. In the subsequent epoch,
the apparent decline in the iPTF14dpk light curve is not statistically significant.

At first blush, the pre-discovery 5-$\sigma$ upper limits of iPTF14dpk in the magnitude plot may mislead the readers into concluding that
the upper limits of iPTF14dpk are roughly consistent with the early detections of iPTF14atg. However, 
when plotted in $\nu L_\nu$, the early detections of iPTF14atg are distinct from the non-detections of
iPTF14dpk at similar epochs by more than 3 $\sigma$
(see the inset of Figure \ref{fig:14dpk:lc}). If we approximate measurement noises with Gaussian distributions, 
the probability that the iPTF14dpk pre-SN non-detections are consistent with the early-phase detections of iPTF14atg is less than $3\times10^{-9}$. 

\subsection{SN-companion interaction}

Next, we seek an explanation to the distinct early-phase light curves of the two otherwise similar SNe. 
Given the spectroscopic typing and almost identical light curves (apart from the early rise), we make a simple and reasonable assumption that 
\textit{all} SN2002es-like events arise from the same progenitor channel. As noted in \citet{ckh+15}, 
the observed early declining UV pulse from 
iPTF14atg provides evidence for the SD progenitor channel. 

Three energy resources may power the optical light curve of a SN from the SD channel: SN shock breakout,
SN-companion collision, and radioactive decay of $^{56}$Ni. The shock breakout of a SN Ia lasts for 
less than a second \citep{pcw10}, so we do not consider this energy resource to explain the iPTF14atg and
iPTF14dpk light curves. 

\begin{figure}
\centering
\includegraphics[width=0.495\textwidth]{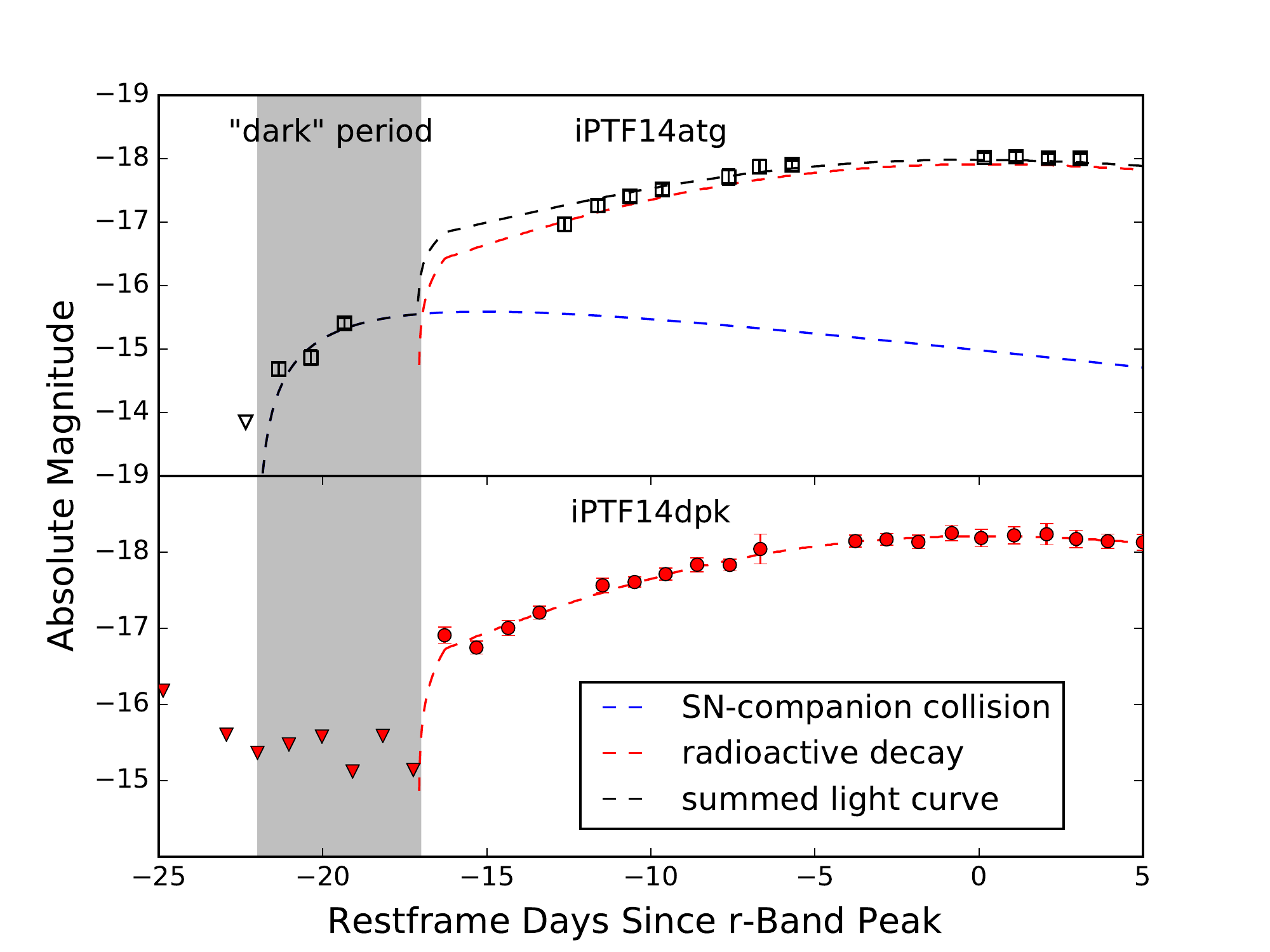}
\caption[Light curve analysis of iPTF14atg and iPTF14dpk]{\textbf{Light curve analysis of iPTF14atg and iPTF14dpk.}
\textit{Upper panel:} The early $R$-band light curve of iPTF14atg has two component: the earlier component (blue dashed curve) 
is the Rayleigh-Jeans tail of the thermal emission computed from SN-companion collision model. The later component
(red dashed curve), powered by $^{56}$Ni, is represented by the iPTF14dpk light curve offset by $0.1\,\textrm{mag}$ to match
the peak magnitude of iPTF14atg. \textit{Lower panel:} The early $R$-band light curve of iPTF14dpk is purely powered by $^{56}$Ni. 
The post-explosion ``dark'' period is highlighted in gray. 
\label{fig:14dpk:14atg}
}
\end{figure}

Both the SN-companion-collision powered and radioactively powered components are seen in the
light curve of iPTF14atg. Specifically, its UV/optical light curve before $-17$ days is dominated by the 
Rayleigh-Jeans tail of the SN-companion interaction signature, and the light curve after $-17$ days is
mainly from radioactive decay. In order to show this, we fit the SN-companion collision model \citep{kasen10} 
to its early UV light curve, Noting that a factor of $\pi$ is missing in the \citet{ckh+15} calculation, 
we find a good fit with an explosion energy of $3\times10^{50}$\,ergs, an ejecta mass of $1.4\sm$, 
and a binary separation of $70\sr$. Then we calculate the \textit{R}-band light curve from this model
and compare it to the observed \textit{R}-band light curve. As shown in the upper panel of Figure
\ref{fig:14dpk:14atg}, the light curve before $-17$ days can be entirely attributed to the Rayleigh-Jeans
tail emission from the SN-companion collision. In later epochs, radiation from the SN-companion interaction
becomes minor and the light curve has to be powered by the radioactive energy. 

In fact, as noted by \citet{kasen10}, given typical mass ratios of a few for SD progenitor binaries of 
SNe Ia, only less than 10\% of the resulting explosions would have geometry that would allow the 
SN-companion collision signature to be seen by a randomly located observer. For the remaining $>90\%$
events, we can only see the radioactively powered light curves. This provides a natural explanation
to iPTF14dpk: the observed iPTF14dpk light curve is purely powered by its radioactive decay. 

If we make a further assumption that both iPTF14atg and iPTF14dpk have similar ejecta structures, then 
the radioactively powered light curves of iPTF14atg and iPTF14dpk have similar shapes. Despite the
observed $0.1$\,mag difference at peak magnitudes which is probably due to the exact amount of 
synthesized $^{56}$Ni, the light curve of iPTF14atg can be roughly treated as a superposition of the 
SN-companion collision light curve and the iPTF14dpk light curve (Figure \ref{fig:14dpk:14atg}).

\subsubsection{A Side Note About iPTF14dpk}
The sharp rise of iPTF14dpk was initially considered to be a possible SN-companion interaction signature, 
but our further analysis soon rejected this hypothesis. According to the scaling relations in \citet{kasen10}, the luminosity from a SN-companion collision 
$L\propto aE^{7/8}M^{-7/8}$ and the effective temperature $T\propto a^{-1/4}$, where $a$ is the binary separation, $E$ is the explosion 
energy and $M$ is the total ejecta mass. Since the temperature is very high, the radiation flux in the $r$ band can be approximated
by the Rayleigh-Jeans law, i.e., $f_\nu\simeq L/(\sigma T^4)\times2kT\nu^2/c^2\propto LT^{-3} \propto E^{7/8}M^{-5/8}a^{1/4}$. 
Hence, the $R$-band flux is insensitive to the binary separation. If iPTF14dpk has an ejecta mass similar to iPTF14atg, then 
in order to fit the first detection of $r=-17\,\textrm{mag}$ of iPTF14dpk, its explosion energy has 
to be $E\gtrapprox2\times10^{51}$\,ergs. Such a large explosion energy would then lead to an expansion velocity significantly
higher than the low expansion velocity of iPTF14dpk derived from its spectral lines. Furthermore, as the effective temperature remains high, 
the expansion of the SN would make the $R$-band flux increase rapidly with time. However, we do not see a fast-rising light curve after the first epoch
of iPTF14dpk. 

\subsection{SN Dark Phase}

Given a binary separation of $70\sr$ and an ejecta velocity of $5\times10^3\,\textrm{km}\,\textrm{s}^{-1}$,
the SN ejecta of iPTF14atg hit its companion star within three hours of SN explosion. 
Here the SN-companion signature provides an accurate approximation to the explosion date for iPTF14atg,
i.e., at $-22.1$ days. Given the similarity between iPTF14atg and iPTF14dpk, it is reasonable to assume 
that the rise time of iPTF14dpk is similar to that of iPTF14atg. Under this framework we explore 
the physical consequences below.

As shown in the lower panel of Figure \ref{fig:14dpk:14atg}, the radioactively powered light curve of iPTF14dpk 
was not visible until $-16$ days. This soon leads to
an interesting result that iPTF14dpk had a ``dark'' period between its explosion and the first light of
its radioactively powered light curve. In our framework, this ``dark'' period also existed in iPTF14atg, but 
was lighted by the SN-companion interaction signature. If iPTF14atg had been observed in a different viewing
angle, then the SN-companion interaction signature would become invisible and hence the ``dark'' period would
appear. 

\begin{figure}
\includegraphics[width=0.495\textwidth]{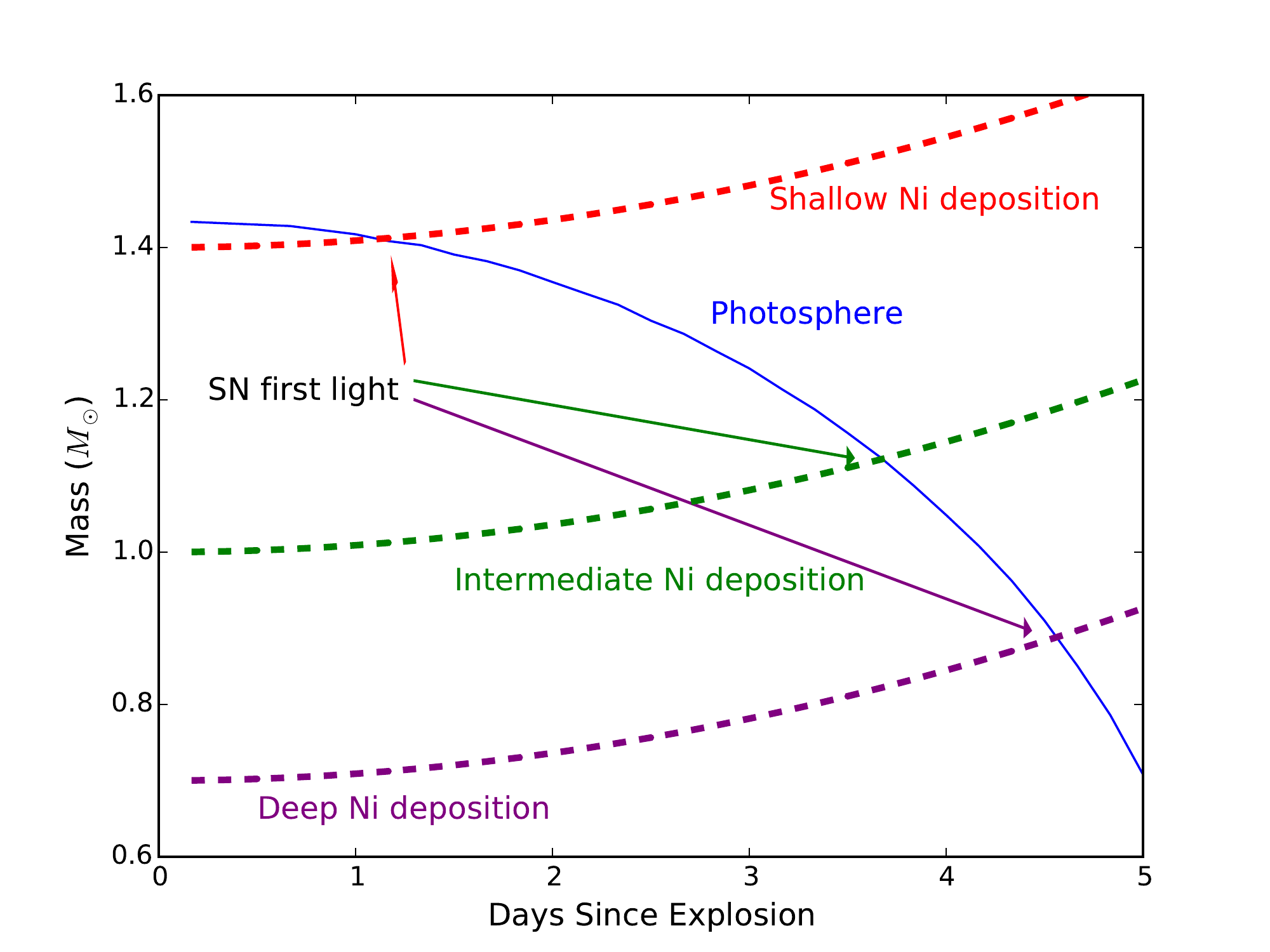}
\caption[Locations of photosphere and diffusive fronts as a function of time since explosion]
{\textbf{Locations of photosphere and diffusive fronts as a function of time since explosion.} 
The $y$-axis is the mass coordinate of the SN ejecta. The location of photosphere (blue solid curve) is calculated for a $1.44\sm$ white 
dwarf model with a constant opacity $0.2\,\textrm{cm}^2\,\textrm{g}^{-1}$ and a SN expansion velocity $10^5\,\textrm{km}\,\textrm{s}^{-1}$. 
The dashed red, green and purple curves correspond to diffusive wavefronts for the shallowest $^{56}$Ni layer located at $1.4\sm$, 
$1.0\sm$ and $0.7\sm$. The propagation of the diffusive wavefront is approximated by $\Delta M\propto t^2$. The intersection points 
between the solid curve and the dash curves represent the time of SN first light. 
\label{fig:14dpk:deposition}
}
\end{figure}

The ``dark'' period is essentially the timescale for the radioactive decay energy to reach the SN
photosphere. During the expansion of the SN ejecta, in the Lagrangian point of view, the photosphere 
moves inwards while the energy diffusive front from the shallowest layer of $^{56}$Ni moves outwards. 
As shown in a toy model in Figure \ref{fig:14dpk:deposition}, a deep deposition of $^{56}$Ni will 
delay the first light of a SN by a few days, compared to a shallow deposition. 

A ``dark'' phase has also been suggested in
the normal Type Ia SN2011fe \citep{hms+13,pn14}. Very recently, \citet{pm15} performed similar but more
sophisticated calculations on the durations of the ``dark'' period for normal Type Ia SNe
with different $^{56}$Ni depositions. They used more realistic WD models from MESA \citep{pbd+11}, added
$^{56}$Ni at different deposition depths, 
and performed radiative transfer calculations with SNEC \citep{mpr+15}. They found that deep deposition
of $^{56}$Ni in a normal Type Ia SN may lead to a ``dark'' period of a couple of days. 

\section{Discussion and conclusion}
\label{sec:14dpk:conclusion}
In this letter, we present optical observations of two low-velocity Type Ia supernovae, iPTF14atg and iPTF14dpk, 
from the intermediate Palomar Transient Factory.
Both are photometrically and spectroscopically similar to the prototypical SN2002es \citep{glf+12}.
Despite their similarities around and after maximum, we observed different initial rise behaviors in these two events. 
While iPTF14atg experiences a steady and slow rise which lasts for 22 days, iPTF14dpk shows a sharp initial 
rise to $-16.9$\,mag within a day of discovery and then joins the rise behavior of iPTF14atg. The apparent
rise time of iPTF14dpk is only 16 days. 

Based on the spectroscopic typing and similar photometric evolution, and based on the observed early declining ultraviolet
pulse in iPTF14atg \citep{ckh+15}, we make a reasonable assumption that all SN2002es-like events arise from the 
single-degenerate progenitor channel. By analyzing the early \textit{R}-band light curves of iPTF14atg and iPTF14dpk, we show that: 
\begin{itemize}
\item The light curve of iPTF14atg can be decomposed into the early SN-companion interaction component and
the late radioactively powered component. The latter component resembles the full light curve of iPTF14dpk.
The absence of the early component in iPTF14dpk is due to an unfavored viewing angle along which the
SN-companion interaction is blocked by the optically thick ejecta. 
\item An SN2002es-like event, or for that matter any Type Ia supernova in general \citep{hms+13,pn14}, 
may experience a ``dark'' phase between the actual explosion and the first
light of its radioactively powered light curve. The duration of the ``dark'' phase is the timescale for
the radioactive decay energy to reach the SN photosphere. 
\item In the case of iPTF14atg, the ``dark'' phase was lighted by Rayleigh-Jeans 
emission from the SN-companion collision.
\end{itemize}

Moving forward, collecting a large sample of young SN2002es-like light curves will be valuable to verify the viewing geometry effect as well as the ``dark''
periods. This sample will also put constraints on the geometry of the progenitor binaries for SN2002es-like events
and thus the mass ratio. For example, if the mass ratio between the donor and the primary white dwarf is one, then
we would have to find 14 young SN2002es-like events, before we could observe the SN-companion interaction again.

The upcoming Zwicky Transient Facility (ZTF; \citealt{sdb+14}) and the planned wide-field fast-cadence ultraviolet surveys,
such as ULTRASAT \citep{sgo+14}, will provide opportunities to collect the sample in both the optical and
ultraviolet, respectively. Assuming that the SN2002es-like event rate is about $10\%$ of the normal Type Ia supernova
rate, we estimate that ZTF will discover eight young SN2002es-like events every year and that ULTRASAT will 
find one dozen in its two-year mission lifetime. These two projects, together with other ongoing and planned
fast-cadence surveys, will provide a big sample to estimate the observational occurrence of the SN-companion
interaction and therefore to put constraints on the geometry of the progenitor binaries. 
Additionally, since the presence of a companion star in a SN explosion 
adds asphericity of the SN ejecta, spectropolarimetric follow-up observations of SN2002es-like supernovae 
will also provide independent constraints on the viewing angles \citep{knt+04}. 

\acknowledgements
YC and PEN acknowledge support from the DOE
under grant DE-AC02-05CH11231, Analytical Modeling for
Extreme-Scale Computing Environments. 
YC also acknowledges support from the National
Science Foundation PIRE program grant 1545949. 
AG is supported by the EU/FP7 via ERC grant no. 307260, the Quantum Universe I-Core programme by the Israeli Committee 
for Planning and Budgeting and the ISF; by Minerva and ISF grants; by the Weizmann-UK `making connections' programme; and 
by Kimmel and ARCHES awards.

This research used resources of the National Energy Research Scientific Computing Center, a DOE Office of Science User Facility 
supported by the Office of Science of the U.S. Department of Energy under Contract No. DE-AC02-05CH11231.
Part of this research was carried out at the Jet Propulsion Laboratory, California Institute of Technology, under a contract with the National Aeronautics and Space Administration.


\end{document}